\documentstyle[aps,prl]{revtex}

\begin{document}
\draft
%
\title{Finite Size Analysis of Luttinger Liquids with a Source of $2k_f$
Scattering}
\author{G. G\'omez-Santos}
\address{
   Departamento de F\'{\i}sica de la Materia Condensada.
   Universidad Aut\'onoma de Madrid, 28049 Madrid, Spain.}
%
%
\maketitle
\begin{abstract}
Numerical analysis  of the
spectrum of large finite size  Luttinger liquids ($g<1$) in the
presence of a single source
of $2k_f$ scattering has been made possible thanks to an effective
integration of high degrees of freedom.
Presence of irrelevant operators and
their manifestation in transport are issues treated independently.
We confirm
the existence of two irrelevant operators: particle hopping and charge
oscillations, with regions of dominance separated by $g=1/2$.
Temperature
dependence of  conductance is shown to be controlled by
hopping alone, giving  $G \sim T^{2/g-2}$ .
Frecuency dependence is affected by both operators,
giving $G \sim \omega^{2/g-2}$ for $g>1/2$, and
$G \sim \omega^{2}$ for $g<1/2$.
\end{abstract}
%

The search for metallic  non-Fermi liquid behavior, motivated by
normal state properties of high $T_c$ superconductors \cite{and}, has
renewed the interest in
one-dimensional (1d) interacting system. Interactions endow
particles in 1d with
properties that are qualitatively different from the quasi-particle
Fermi liquid picture, requiring a new description under the name  of
Luttinger liquids (LL) \cite{hal}. Transport is severely
affected and, for instance,
the conductance of a perfect channel passes from being universal to
depend on the interaction strength \cite{kf}. More striking is the
effect of even a single impurity:
depending on the host conductance, the
imperfection can either transmit or reflect perfectly \cite{kf}.
Although there
have been great advances in the nanofabrication of 1d structures
\cite{timp}, the
realization that edge states of quantum Hall effect liquids are chiral
LL \cite{wen} has opened new perspectives to this field. Now it seems
possible to give experimental reality \cite{exp} to
idealizations such as a single source of $2k_f$ scaterring in an
otherwise perfect LL.

A LL in the presence of a localized source of $2k_f$ scattering is
described by \cite{hal,kf}
\begin{equation}
\label{H}
H=\int dx \frac{1}{2} (\pi g \Pi(x)^2 + \frac{1}{\pi g}(\partial_x
\theta)^2 - u \delta(x) \cos(2\theta))
\end{equation}
where $\theta$ and $\Pi $ are canonical conjugate fields  related to
charge
and current respectively, $g$ is the conductance in the absence of
perturbation, and the impurity scattering is given by the last cosine
term.
For $ g<1 $ (repulsive case)
the perturbation is  relevant \cite{guin1,kf},
effectively splitting the system into two parts, with the phase
at the impurity location, $\theta_o=\theta(x=0) $,
pinned in any of the equivalent minima of the
cosine term. The
pinning of $\theta_o $ blocks transport driving the
conductance to zero \cite{kf}. At low energy,  $\theta_o $ owes its
residual dynamics
to  hopping processes between cosine minima (particle transport
across the junction). This hopping is an
irrelevant perturbation with scaling dimension $1/g$, giving
the current-current correlator at the junction the
imaginary time ($\tau$) dependence $\tau^{-2/g}$. Standard scaling
for Kubo formula would then provide the following temperature ($T$)
or frecuency ($\omega$) dependence to the conductance: $G \sim
\eta^{2/g-2} $, with $ \eta=\omega \; or \;  T $, as Kane and Fisher
(KF) have shown \cite{kf}.

 Based on the Self-Consistent Harmonic Approximation (SCHA)
\cite{zwe,go}, which
substitutes the effect of the cosine in Eq. (\ref{H}) by a quadratic
term $ \alpha \; \theta_o^2 $, Guinea {\it et al.} \cite{guin} have
qualified this picture. They  claim that, in addition to the
hopping between minima, the SCHA suggests the existence of another
irrelevant perturbation of dimension $2$, describing oscillations of
$\theta_o$ around its pinned value and providing the junction with a
capacitance.
(A density-density coupling between  both sides of the  already split
system would be an equivalent description.)
This new irrelevant operator (dominant for $g<1/2$) would
give rise to a $g$-independent $\tau^{-4}$  contribution to the junction
current-current correlation. Scaling then dictates a $\omega^2$
dependence for
the conductance,  replacing   the KF result
$\omega^{2/g-2}$ when $ g<1/2 $.

The case for this new operator seems strong. Based on a perturbative
analysis, Chamon {\it et al.} \cite{cham1} have found
evidence for the $\tau^{-4}$ dependence of the current-current
correlator for $g<1/2$. The same
authors \cite{cham2} have been forced to invoke the presence of a
density-density
coupling to understand features of the noise spectrum, and Tsvelik
\cite{tsv} has deduced the existence of a capacitance from  Bethe ansatz
techniques.
On the other hand, the same scaling that gives
the $ \omega^2$ law for the dynamic conductance
would predict  a $ T^2 $ temperature dependence for the
static conductance \cite{guin} replacing the KF prediction for $g<1/2$.
Unfortunately this last result contradicts
the exact solution found for $g=1/3$ from Bethe ansatz techniques
\cite{fen}, which is in agreement with
the KF analysis $G \sim T^{2/g-2}$. This confronts us with a puzzling
situation.

  Naive scaling is not guaranteed to work,
and to settle the problem described  above one
would like to  separate the following two issues: (i) presence of
irrelevant operators (nature,
dimension, and region of dominance), and (ii) their manifestation in
physical properties. With this in mind, we have adopted a finite size
approach in this paper.
The idea is well known: solving $H$
in a finite size $L$ amounts to probing the system  with a
energy scale  set by
$ 1/L$.
The low energy spectrum of the
(scaled)
Hamiltonian $L H  $ becomes scale invariant in the limit $L=\infty$
(fixed point).
The presence of irrelevant operators and their scaling dimensions
can be read off from
the $L$ depedence of the approach to the fixed point.

 Unfortunately, this simple idea faces severe
implementation problems. The number of degrees of freedom grows
gigantically with size, and there is no real chance of
solving exactly
a finite system of even modest size.
One should resort to approximations that keep a manageable number of
states while ensuring that they  truly describe the
low energy sector of interest.
In our case, the use of
the SCHA as an intermediate step will help us handle this problem in a
transparent and reliable way, as we will see.

 It is well known that, under the guise of a variational approach,
the SCHA
performs a perturbative renormalization \cite{zwe,go,guin}: replacement
of the cosine
term in Eq. (\ref{H}) by a self-consistent $\alpha \theta_0^2$  provides
the following dependence: $\alpha \sim u^{1/(1-g)} $. This behavior
is precisely what would have been obtained for the running
coupling $u$ upon integration of its renormalization group (RG) flow up
to the point where $u$ equals the cut-off energy, signalling the
end of the perturbative regime. Thus, the low energy states of the SCHA
have a
built-in effect of the high energy modes. Although the SCHA flows to the
same fixed point expected for the original problem ( $\theta_o$
is also pinned), it loses
memory
of the discrete (global) translational invariance ( $\theta \rightarrow
\theta+n\pi$) of $H$. It is clear that
the SCHA can pin $\theta_o$ at  any of the translation-equivalent
minima, suggesting the
following procedure: replicate the SCHA states for each  well,
and use these states as a basis for the
diagonalization of the original Hamiltonian.

The steps in the precise
implementation of  this idea are the following.
(i) The Hamiltonian $H$ is regularized in a linear chain of $L$ sites
with periodic boundary conditions, the cosine perturbation
affecting only one lattice site.
(ii) The SCHA is performed for every $L$, providing a basis which is
replicated at every equivalent well.
(iii) The regularized hamiltonian is scaled, $h(L)= H L/\pi $, and
diagonalized
in the replicated  SCHA basis, keeping a fixed number of states ($\sim
100$) per well and taking advantage  of the $\theta$ translational
symmetry.

It is important to stress  that
the {\em true} Hamiltonian is diagonalized in a restricted basis
provided by  he SCHA as an intermediate step.
The effective integration of high modes implied by the SCHA will
prove crucial for the success of this approach.
Matrix elements and overlaps pose no severe computational problem
owing to the harmonic nature of the basis.
 The calculation has all the
ingredients of  tight-binding band structure problem,
and the spectrum adopts the form of enery
$(\epsilon)$ versus {\it crystalline} momentum $(q)$, associated with
the (discrete) $\theta$-periodicity.
$\theta$ being charge, $q$ can be identified
with flux piercing the
ring, enabling transport properties to be read off from
$\epsilon(q)$ \cite{kohn}.

 We have performed calculations following the above described procedure
for values of g in the interval $ 1/3 \leq g < 1$, with lattice sizes
in the range $ 100 < L < 10^7 $,
and values of u bounded by $u < 10^{-1}$ (much less than the energy
cut-off). The
number of states per $\theta_o$ well is $100$, having checked that
doubling
this number produces no significant changes. The first important point
to remark is  the $ (u,L)$ dependence of the spectrum of the scaled
$h(L)$
through the combination $u^{1/(1-g)}L$, allowing us to measure the
effect of the perturbation by the following scaling variable:
\begin{equation}
\label{utilde}
     \tilde{u}(u,L) = u (2/L)^{g-1}
\end{equation}
 Trading $1/L$ for temperature, this is the same scaling variable found
by KF \cite{kf} to describe their conductance results. This scaling is a
consecuence of the RG
flow for the relevant coupling $u$. In our scheme, this property
comes through the mentioned dependence $ \alpha \sim u^{1/(1-g)} $ of
the SCHA,
emphasizing again its implicit integration of high degrees of freedom.

 Representative results for the spectrum are shown in Fig. 1, where
the two lower bands $\epsilon_{1,2}(q)$ of $ h(L) $ for $g=0.4$ are
plotted for increasing
values of $\tilde{u}$. (Similar results are obtained for all values of
$g$.)
The spectrum evolves smoothly from the fixed point $ \tilde{u}=0^+$
\cite{ucero}
to that of $\tilde{u}\rightarrow\infty$, where (flat band) energies
and degeneracies are, of
course, those of a broken chain. Although our interest is the approach
to this last fixed point, the analysis of the departure point $\tilde{u}
\sim 0$
provides a critical test of our treatment. In
the left panel of Fig. 1 we have superimposed the exact result for the
(folded) parabola of current-carrying states of a perfect LL with our
results. The agreement is perfect, and one should emphasize that the
scheme of our calculation is the same for all values of $\tilde{u}$.
That means, in the solid-state language, that we are reproducing the
free particle dispersion starting out of a tight-binding basis.
Following this parabola upwards in energy,
discrepancies between exact and calculated results appear,
due to the limited basis employed (this happens around $\epsilon \sim
5$ in the present case). This provides us with a quantitative measure of
the validity of our results. All calculations presented here are
for energies within
this ($\tilde{u}=0$) confidence limit, even though it is clear
that this energy window will widens with increasing $\tilde{u}$.
A further point to notice in the small $\tilde{u}$ limit is that the
{\em Bragg}
gap, $\delta \epsilon_{B}\equiv\epsilon_2(\pi)-\epsilon_1(\pi) $, opens
with a dependence $\delta\epsilon_B \sim \tilde{u}\sim L^{1-g}$. This is
the expected perturbative
behavior for  the  $2k_f$ perturbation  of a LL, which grows with
scaling dimension $g$.
This agreement in the nominally
worse situation for a scheme based on a tight-binding approach,
gives us confidence in the quality of our approach.

 Now we study irrelevant operators. The most conspicuous effect in the
approach to the $\tilde{u}=\infty $ fixed point
is the lack of band dispersion. This reflects the vanishing particle
hopping across the junction (tunnel between $\theta_o$ minima). We
characterize this residual hopping
by the dispersion of the lowest band, and present
its $L$ dependence in Fig.2 (left panel). For
all values of $g$ it shows the dependence
\begin{equation}
\label{e1}
\delta \epsilon_1 \equiv (\epsilon_1(\pi)-\epsilon_1(0)) \sim L^{1-1/g},
\end{equation}
corresponding to a scaling dimension $1/g$, in agreement with the
KF analysis for the hopping across the junction.

Let us now consider other clear feature of the spectrum in Fig. 1: the
approach of the second band to its asymptotic value at the
Brillouin zone center,
$\delta \epsilon_2\equiv (1-\epsilon_2(0))$.
Its $L$ dependence is plotted in Fig. 2
(right panel) and summarized as follows:
\begin{equation}
\label{e2}
\delta \epsilon_2 \equiv |1-\epsilon_2(0)| \sim \left \{
\begin{array}{ll}
L^{1-1/g} & (1/2 < g < 1) \\
L^{-1}    & (g < 1/2)
\end{array}
\right.
\end{equation}
If tunnel between wells
were the only irrelevant
operator, then we would always have the first result of Eq. (\ref{e2})
This is certainly what happens for
$g>1/2$, while for $g<1/2$ the behavior obtained is precisely that
expected
from the presence of an operator with scaling dimension $2$
describing
fluctuations of the pinned phase $\theta_o $, as explained before.
These results confirm the picture for  two
irrelevant operators described at the beginning, with the value $g=1/2$
marking the boundary between regions of dominance.

Now we study the
consequences for transport properties, beginning with the
temperature dependence of the static conductance.
Although we cannot calculate
the conductance of the {\em infinite} system as a function of $T$,
we can circumvent this problem by defining a running
temperature for each size, $T=2/L$, so that the energy scale for a
given size  is also the temperature. We {\em define} the conductance
for each size-temperature according to
\begin{equation}
\label{cond}
\tilde{G}(u,L,T=2/L) = \pi^2 < \partial ^2_q \epsilon >_T,
\end{equation}
where its well known expression as a thermal average ($<>_T$) of the
second derivative of energy versus flux has
been used \cite{kohn}.

Although the $\tilde{G}$ so calculated  is not the true (infinite size)
conductance as a function of $T$,
 it is clear that it will be
a universal function of the scaling variable $\tilde{u}$ [Eq.
(\ref{utilde})],
 with  the same scaling dependences.
Results for $\tilde{G}(\tilde{u})$ are plotted in Fig. 3 (left
panel) for $g=0.4$. Notice the correct limit
$\tilde{G}(\tilde{u}\rightarrow 0) \rightarrow g$, and a smooth
crossover to the asymptotic behavior
$\tilde{G}(\tilde{u}\rightarrow \infty) \rightarrow \tilde{u}^{-2/g}$.
This limit implies (remenber $T=2/L$) that $\tilde{G} \sim T^{2/g-2}$,
in agreement with the KF result.
Similar results have been obtained for  all values
of $g$, at both sides of $g=1/2$. This should not surprise in view of
Eq. (\ref{cond}), where only the operator responsible for
band dispersion can appear.

Now we study the dynamic conductance, that is, the $\omega$ dependence
of the (real) part of $G$ (zero temperature). Placing a time
dependent voltage (represented by a vector potential) right at the
junction
adds a term to the $H$ proportional to the current
density $\Pi_o$. Calculating matrix element  of $\Pi_o$,
we can extract the power loss
associated with a transition frecuency, whereupon $G(\omega)$
can be obtained.  We define a (scaled)
{\em oscillator strength} $f_{12}(q)$  for the vertical transition
between states of the two lowest bands as
\begin{equation}
\label{oscil}
f_{12}(q)=(g/2)|<q,2|\Pi_0|q,1>|^2 L^2,
\end{equation}
and analyze  its $L$ dependence for very large sizes. The results are
presented in Fig. 3 (right panel),  where the
(Brillouin-zone averaged)
oscillator strength, $ \bar{f}_{12}=<f_{12}(q)>_q $, is plotted
for several values of $g$. The following
asymptotic behaviors are obtained:
\begin{equation}
\label{o}
\bar{f}_{12} \equiv <f_{12}(q)>_q \sim \left \{
\begin{array}{ll}
L^{2-2/g} & (1/2 < g < 1) \\
L^{-2}    & (g < 1/2)
\end{array}
\right.
\end{equation}
It is a simple exercise to show that $\bar{f}_{12} \sim L^{-\eta}$
implies $G(\omega)\sim \omega^{\eta}$.
Thus, our results show that the dynamic
conductance changes from $ G(\omega)\sim \omega^{2/g-2}$ for
$1/2<g<1$, to $G(\omega)\sim \omega^2 $ for $g<1/2$.

This is in  agreement with the picture presented in the
introduction,
meaning that both irrelevant operators contribute
to $G(\omega)$. For $g>1/2$ hopping dominates and one obtaines the KF
result.
For $g<1/2$ oscillations of the pinned $\theta_o$ dominate, and
$G(\omega)$ reflects
the capacitance interpretation of this irrelevant peturbation. It is
interesting to notice that a simplified analytical treatment that keeps
only the two lowest states per well, shows that the oscillator
strength comes from two contributions. One  performs the
transition for the  states
of the same well (no particle tunnelling), and is dominated by
the {\it capacitance} operator. The other makes the transition between
states of different  wells, $\theta_o \rightarrow \theta_o \pm \pi$, and
only depends on the hopping irrelevant operator. This
difference would help explain why the capacitance operator does not show
up in the static conductance: its contribution to the oscillator
strength is through transitions conserving $\theta_o$ (charge).

In summary, we have studied the low energy properties of a LL in the
presence of $2k_f$ scattering in the repulsive regime $g<1$.
The numerical analysis  of the
spectrum for very large sizes has been made possible thanks to the
use
of the SCHA as an intermediate step that effectively integrates high
energy states. Accuracy has
been tested in the nominally worst case of small $u$. Presence
of irrelevant operators and
their manifestation in transport are two issues  treated
independently. We confirm
the existence of two irrelevant operators, with $g=1/2 $ as the
boundary between their regions of dominance. One represents
particle hopping,
and has dimension $1/g$. The other describes charge oscillations
around the pinned value, and has dimension $2$. Temperature
dependence of the static conductance has been show to be controlled by
the hopping operator alone, giving  $G \sim T^{2/g-2}$ for all values
of $g$.
Frecuency dependence is affected by both operators,
giving $G \sim \omega^{\eta}$, with $\eta=(2/g-2) $ for $g>1/2$, and
$\eta=2 $ for $g<1/2$. Conceptual simplicity, tested numerical accuracy,
and versatility in handling different aspects make this approach a
promising tool for other quantum impurity problems.

The author is grateful to Prof. F. Guinea for very stimulating
discussions.
This work has been supported by the DGICyT of Spain (grant PB92-0169).

%
%

%
%
%
\begin{figure}
\caption{Two lowest bands $\epsilon_{1,2}(q)$ of the scaled
hamiltonian $h(L)=H L/\pi$
with $g=0.4$ for four increasing values of the scaled perturbation
[Eq.(2)]
$\tilde{u}=0^+,0.72,1.63$   and   $34.9$ (left to right).
The continuous line of left panel is the exact result
of a perfect LL.}
\end{figure}
\begin{figure}
\caption{Left panel: $ln(\delta\epsilon_1)$ [Eq.(3)] versus
$ln(L/\pi)$ for $g=0.7,0.55,1/2,0.4,$ and $1/3$ (top to bottom).
Continuous lines are asymptotic behavior [Eq.(3)].
Right  panel: as in left panel for $ln(\delta\epsilon_2)$ [Eq.(4)]
with $g=0.7,0.55,1/3,0.4,$ and $1/2$ (top to bottom).}
\end{figure}
\begin{figure}
\caption{Left panel: $ ln(\tilde{G}) $ [Eq.(5)] versus
$ ln(\tilde{u})$ [Eq. (2)] for $g=0.4$ with results for 200
$(u,L)$ points collapsed onto a single scaling curve.
Continuous line is  the asymptotic behavior $\sim \tilde{u}^{-2/g}$.
Arrow marks the exact limit for $\tilde{u}=0$.
Right  panel: $ln(\bar{f}_{12})$ [Eq.(7)] versus
$ln(L/\pi)$ for $g=0.7,0.55,1/2,1/3,$  and  $0.4$ (top to bottom).
Continuous lines are asymptotic behavior [Eqs.(7)].}
\end{figure}

\begin{references}
%
\bibitem{and} P. W. Anderson, Phys. Rev. Lett. {\bf 64}, 1839 (1990).
%
\bibitem{hal} F. D. M. Haldane, J. Phys. C {\bf 14}, 2585 (1981).
%
\bibitem{kf} C. L. Kane and M. P. A. Fisher, Phys. Rev. Lett. {\bf 68},
1220 (1992); Phys. Rev. B {\bf 46}, 15233 (1992).
%
\bibitem{timp} G. Timp in {\it Mesoscopic Phenomena in Solids}
edited by B. L. Altshuler, P. A. Lee, and R. A. Webb (Elsevier,
Amsterdam, 1990).
%
\bibitem{wen} X. G. Wen, Int. J. Mod. Phys. B {\bf 6}, 1711 (1992).
%
\bibitem{exp}  F. P. Milliken, C. P. Umbach, and R. A. Webb, preprint.
%
\bibitem{guin1} F. Guinea, V. Hakim, and A. Muramatsu,
Phys. Rev. Lett. {\bf 54}, 263 (1985).
%
\bibitem{zwe} M. P. A. Fisher and W. Zwerger,
Phys. Rev. B {\bf 32}, 6190 (1985).
%
\bibitem{go} A. O. Gogolin, Phys. Rev. Lett. {\bf 71}, 2995 (1993).
%
\bibitem{guin} F. Guinea, G. G\'omez-Santos, M. Sassetti, and M. Ueda,
Europhys. Lett. {\bf 30}, 561 (1995).
%
\bibitem{cham1} C. de C. Chamon, D. E. Freed, and X. G. Wen, Phys. Rev.
B. {\bf 51}, 2363 (1995).
%
\bibitem{cham2} C. de C. Chamon, D. E. Freed, and X. G. Wen,
preprint cond-mat/9507064.
%
\bibitem{tsv} A. M. Tsvelik,
preprint cond-mat/9409027.
%
\bibitem{fen} P. Fendley, A. W. W. Ludwig, and H. Saleur, Phys. Rev.
Lett. {\bf 74}, 3005 (1995); preprint cond-mat/9503172.
%
\bibitem{kohn} W. Kohn, Phys. Rev.  {\bf 133}, 171 (1964).
%
\bibitem{ucero} The calculation requires a minimum value of $u \ll 1/L$
to be technically possible. For all practical purposes this means
$\tilde{u}=0$.
%
\end{references}
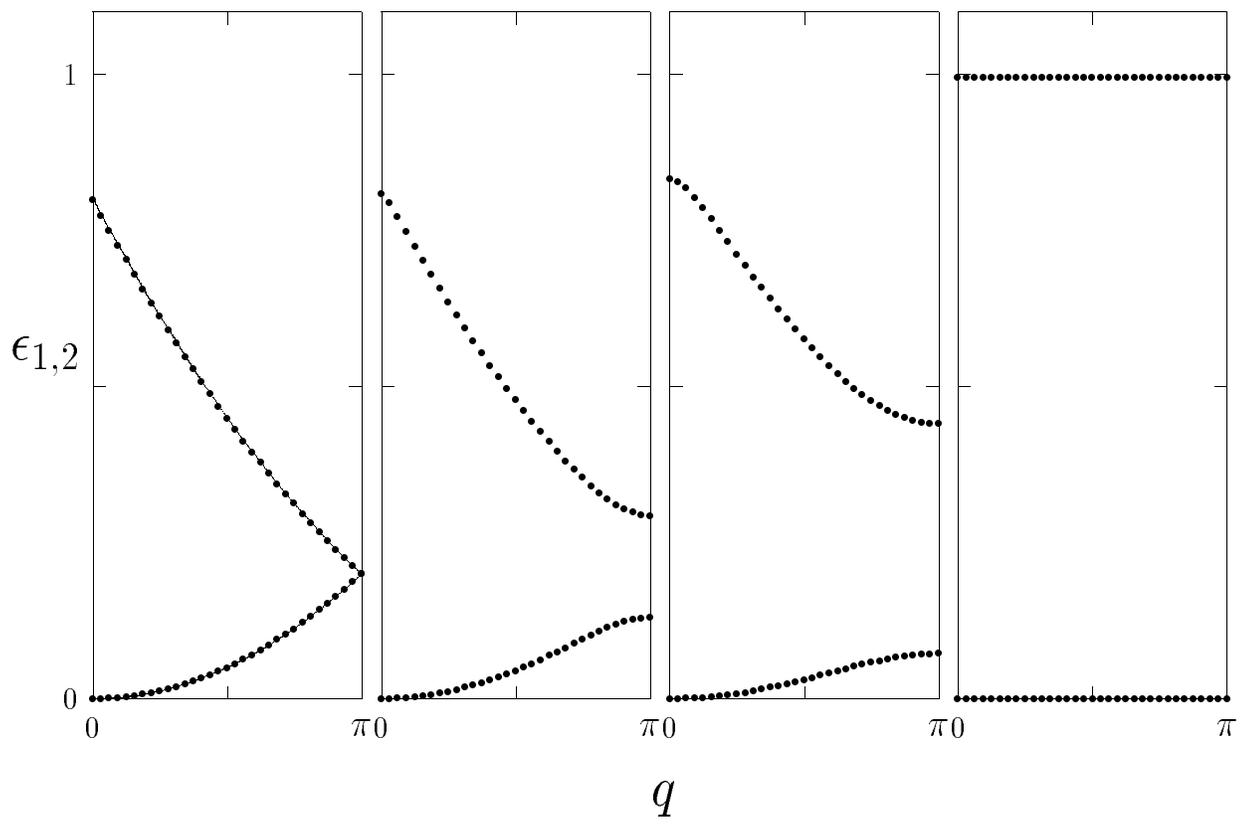
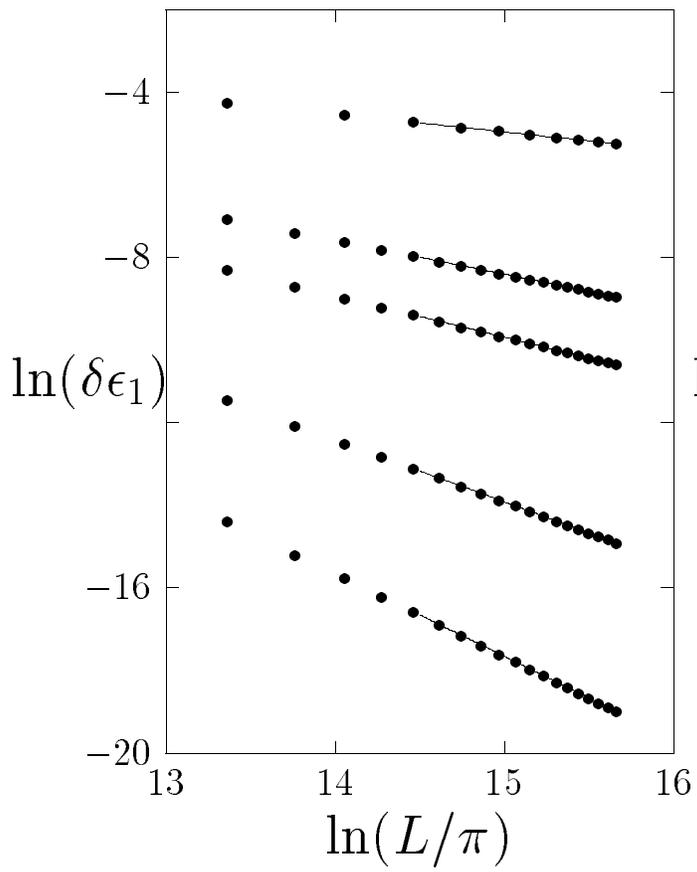
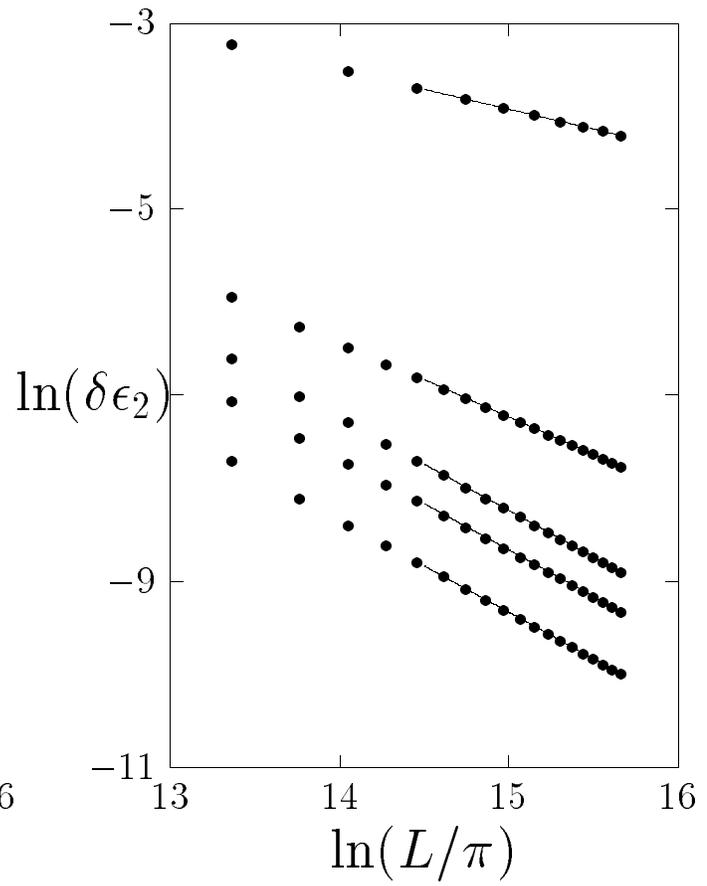
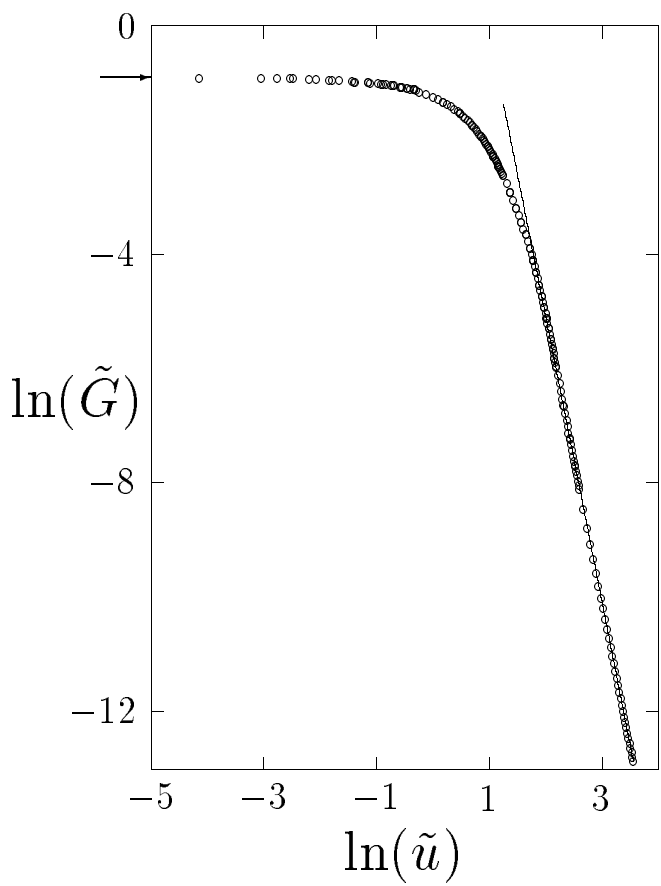
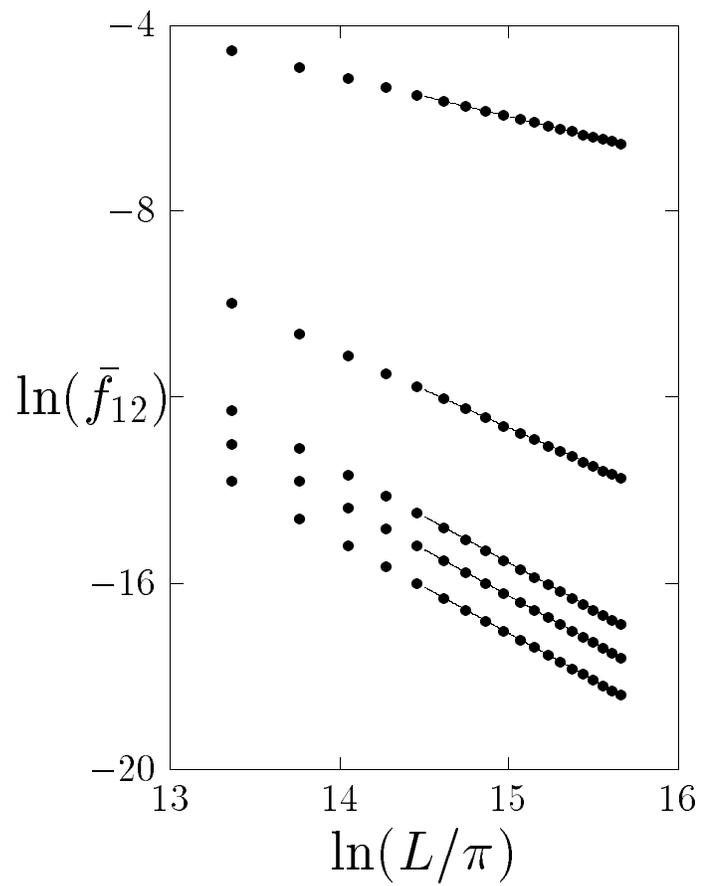
\end{document}